\documentclass[12pt]{article}

\usepackage{amssymb}
\usepackage{graphics}
\usepackage{citesort}

\newcommand{\beq}{\begin{equation}}
\newcommand{\eeq}{\end{equation}}

\begin{document}

\begin{titlepage}

\pagestyle{empty}

\setlength{\baselineskip}{18pt}

\rightline{hep-ph/99mmnnn}
\rightline{Alberta Thy-XX-99}
\rightline{CERN-TH-99-182}
\rightline{July 1999}

\vskip .2in

\begin{center}

{\large{\bf Parametric Resonance For Complex Fields}}
\end{center}

\vskip .1in

\begin{center}
Rouzbeh Allahverdi and R. H. A. David Shaw 

{\it Department of Physics, University of Alberta}

{\it  Edmonton, Alberta, Canada T6G 2J1}

Bruce Campbell

{\it Theory Division, CERN}

{\it CH1211 , Geneve 23, Switzerland}

\vskip .1in

\end{center}

\centerline{ {\bf Abstract} }

\setlength{\baselineskip}{18pt}

\noindent

Recently, there have been studies of parametric resonance decay of 
oscillating real homogeneous cosmological scalar fields, in both the 
narrow-band and broad-band case, primarily within the context of 
inflaton decay and (p)reheating. However, many realistic models of 
particle cosmology, such as supersymmetric ones, inherently involve 
complex scalar fields. In the oscillations of complex scalars, a 
relative phase between the oscillations in the real and imaginary 
components may prevent the violations of adiabaticity that have been 
argued to underly broad-band parametric resonance. In this paper, we 
give a treatment of parametric resonance for the decay of homogeneous 
complex scalar fields, analyzing properties of the resonance in the
presence of out of phase oscillations of the real and imaginary 
components. For phase-invariant coupling of the driving parameter 
field to the decay field, and Mathieu type resonance, we give an 
explicit mapping from the complex resonance case to an equivalent 
real case with shifted resonance parameters. In addition, we consider
the consequences of the complex field case as they apply to ``instant
preheating,'' the explosive decay of non-convex potentials, and
resonance in an expanding FRW universe.  Applications of 
our considerations to supersymmetric cosmological models will be 
presented elsewhere.

\vskip .1in

\centerline{{Submitted For Publication In: {\it Physics Letters B}}}

\end{titlepage}

\setlength{\baselineskip}{18pt}



\section{Introduction}

With the advent of inflationary theories of the early universe, it 
has been argued that the present stage of hot FRW ``big-bang'' 
cosmology was preceded by an epoch of cosmological evolution 
dominated by the dynamics of scalar fields \cite{infl}.  The success 
of inflationary models in providing explanations for flat large-scale 
geometry (as suggested by the location of the acoustic peaks in the 
CMBR anisotropies), and for the origin of approximately scale-free 
adiabatic density perturbations (which can be used to simultaneously 
fit both the CMBR anisotropies and observations of cosmic structure 
formation), lends support to the idea of an early scalar-field 
dominated epoch. A crucial question in this picture is the nature of 
the transition from the scalar field dominated epoch, to the hot FRW 
epoch, which is referred to as reheating. The nature of this 
transition also relates to other aspects of early universe dynamics 
necessary for a successful cosmology, such as mechanisms of 
baryogenesis, the resolution of cosmological moduli problems, and 
possible sources for non-thermal dark matter.

The standard approach to reheating, which applies to sufficiently  
weakly coupled inflaton fields \cite{sr}, is to treat quanta of the 
inflaton field as particles, which undergo independent single 
particle decay; this treatment, if adequate, has the advantage that 
the post-inflation reheat temperature is determined by the 
microphysics of the model. For inflaton fields with mass as suggested 
by the simplest chaotic or supersymmetric models, and decaying by 
gravitational strength interactions, this treatment is adequate,
leading to moderate reheat temperatures 
($\lesssim \mathrm{O}(10^{10})$ GeV) which are consistent with the 
absence of GUT-scale defects such as monopoles, which are capable of 
incorporating a variety of (s)leptogenesis or electroweak mechanisms 
for generation of the BAU, and which avoid, in the supersymmetric
case, cosmological problems with thermal overproduction of gravitinos 
after reheating \cite{ekn}.

Recently it has been realized that the standard treatment of reheat 
in terms of single particle inflaton decay may be seriously 
misleading in circumstances where there is coherent enhancement of 
the transition to bosonic decay products
\cite{kls1,kls2,stb,k1,kt1,kt2,pr,gkls/k2,gpr,zhs,gk,gks,fkl}.
For large mode occupation numbers of the decay product field we may  
treat its dynamics as being essentially classical. Mode by mode for 
the decay product field its coupling to the oscillating inflaton 
field induces a periodic time dependence in the mode mass (modulated 
by cosmic expansion which ``sweeps'' the time dependence of each 
comoving mode through the bands of the stability chart of the mode 
equations.) This periodic modulation of the parameters of the 
oscillator associated with each mode of the decay product field, can 
induce parametric resonance in bands of the mode parameters, leading 
to exponential growth in the decay mode amplitude.  The resonant 
decay of the inflaton may have important cosmological implications 
like non-thermal symmetry restoration and subsequent formation of 
topological defects \cite{nsr,tf}, revival of GUT baryogenesis 
scenarios \cite{gb1,gb2}, supersymmetry breaking \cite{sb}, 
superheavy particle production \cite{fkl,spp}, and gravitino 
production \cite{gp1,gp2}. 

The exponential growth in the mode occupation number may be modified
or regulated by a number of physical processes. These include the 
decay of produced quanta to other particles \cite{gb2,kk} or the 
rescattering of final state particles \cite{kt2}. Another possibility,
occurring in models with gauge-strength self-interactions between the 
produced final state particles, is the regulation of the parametric  
resonance by the self-interaction induced effective masses of the  
produced quanta, which can move these quanta out of the available  
resonance bands; in this case, resonance only proceeds as 
thermalization and Hubble dilution reduce the plasma masses of the 
final state quanta, resulting in a quasi-steady-state resonance 
conversion of inflaton oscillation energy to decay products \cite{ac}.
This general scenario for the physically realistic case of decay 
products with gauge charge has been verified in explicit calculations
in the narrow-band resonance case \cite{ac}, and in numerical 
simulations in the broad-band case \cite{pr}.

While to date analytical and numerical treatments of parametric  
resonance have considered oscillations of a single real field 
decaying to a single real decay product field, in realistic models 
the field content is often more extensive. In the case of 
supersymmetric theories this is unavoidably the case, as the physical 
scalars of simple (N=1) supersymmetry come as components of chiral 
multiplets and are complex. So for these types of theories, we 
should at the very least consider the nature of coherent decays when 
the fields involved are complex, though non-supersymmetric models 
with multiple real scalar fields may share some of the features of 
the simplest complex case.

Within supersymmetric models of particle physics, there are several  
different circumstances under which the decay of a homogeneous complex
scalar condensate may occur in the early universe. At the end of  
inflation one expects to have a spatially homogeneous inflaton scalar  
condensate, whose decay energy will ultimately be responsible for  
cosmic reheating and the initiation of hot big-bang cosmology. As 
well, in the supersymmetric standard model there are directions in 
the scalar field space of squarks and sleptons which are F-flat and 
D-flat, and which only gain a potential from supersymmetry breaking. 
In the early universe these directions may be populated with (very) 
large vev's after the end of the inflationary epoch; these vev's may 
carry enormous vev to mass ratios (Mathieu resonance parameter $q$) 
and couple to other directions in scalar field space with couplings 
capable of inducing resonant decay. Finally, supersymmetric models 
are generically plagued with gauge singlet scalar moduli, whose
homogeneous oscillation poses grave cosmological difficulties which 
might be ameliorated by coherent decay of their oscillation 
amplitude.

For self-interactions of complex scalars of the general form dictated
by the F-term and D-term couplings arising in globally supersymmetric
theories, the fields generally appear in complex conjugate pairs for 
the F-terms and diagonal D-terms. For example, let us consider a 
complex scalar field $\Phi$ in a chiral supermultiplet whose decay 
will be induced by a trilinear (renormalizable) coupling in a 
superpotential $W$ to a chiral multiplet labelled by its scalar 
$\Xi$, where $W \simeq \Phi\Xi\Xi$. The resulting F-term coupling 
inducing the decay is then of the form ${\Phi^{*}}\Phi{\Xi^{*}}\Xi$, 
and is invariant under global phase redefinitions of either the 
$\Phi$ or the $\Xi$. We will see in the next section that in cases 
such as this the phase invariance of the resulting couplings implies 
that the equations for modes of the real and imaginary components of
$\Xi$ are decoupled and independent, and will allow us to simply 
analyze the resonant decay of a $\Phi$ condensate with out of phase 
oscillations for the real and imaginary components of $\Phi$, into 
real and imaginary components of the decay product field $\Xi$. 

We can always phase rotate our scalar field $\Phi$ to a basis such 
that its initial vev lies along the real axis.  If there is no 
component of force along the direction of the imaginary axis 
(i.e. the scalar potential is phase invariant), the trajectory of 
the motion of $\Phi$ is limited to the real axis and the field hits 
the origin as it oscillates back and forth.  In this case, provided 
that the coupling of the oscillating field to the final state field 
is also phase invariant, the situation is exactly that of a real 
oscillating field, and the same arguments apply for parametric 
resonance particle production.  However, if the scalar potential is 
not phase invariant, i.e. depends on the phase of the oscillating 
field as well, a torque is exerted on the field. This leads to the 
deflection of the trajectory from a straight line and results in 
changing the trajectory into something that finally resembles an 
ellipsoid, after the torque in field space has effectively ceased its
action. In this case the field no longer passes through the origin
but rather has a finite distance of closest approach to it.  This
will have important implications for broad-band parametric resonance
as we discuss below.

In supersymmetric models not only are complex scalar fields 
inherently involved, but also a phase dependent part of the scalar 
potential can arise naturally from supersymmetry breaking.  Let us 
consider the simplest case with the following terms only involving
the inflaton in the superpotential 
$W = {1 \over 2} m {\Phi}^{2} + {1 \over 3} \lambda {\Phi}^{3}$.  
In supergravity models with broken supersymmetry, there is a 
corresponding phase dependent term (the ``A-term'') 
$A m_{\Delta} {\partial {W} \over \partial \Phi} + \mathrm{h.c.}$ in 
the scalar potential, where $A$ is a dimensionless model-dependent 
constant and ${m}_{\Delta}$ is the scale of supersymmetry breaking 
in the sector in which $\Phi$ lies.  There is also a phase dependent 
term $m {\lambda}^{*} \Phi {{\Phi}^{*}}^{2} + \mathrm{h.c.}$ in the 
F-term part of the scalar potential.  This generically occurs in 
minimal supergravity models for inflation where the superpotential 
contains a series of ${\lambda}_{n}{{\Phi}^{n} \over {M}^{n-3}}$ 
terms \cite{drt}, and occurs even in ``no scale'' supergravity models
after the inclusion of radiative coreections to the effective
potential \cite{gmo}. 

For $V \supset {\Phi}^{m}{({\Phi}^{*})}^{n}$ the potential along  
the angular direction is periodic with $m-n$ minima.  In general,  
during inflation $\Phi$ rolls down to its minimum both along the 
radial and angular directions.  In order to have a torque to deflect 
the trajectory, $\Phi$ must not be at the minimum along the angular  
direction at the onset of radial motion.  This can happen in two 
ways: either there are several phase dependent parts of the potential 
with a non-adiabatic transition from the minimum of one to another, 
or $\Phi$ does not settle at the minimum along the angular direction.  
The first possibility happens when other supersymmetry breaking 
sources in the early universe (e.g. non-zero energy density of the 
universe or finite temperature effects) are dominant over the low 
energy one.  In this case the minimum in the angular direction at 
early times is different from that at late times (due to independent 
phases for the coefficients of different A-terms).  If the transition
from one minimum to another one is non-adiabatic, $\Phi$ will not be 
at the minimum at late times regardless of its start at the minimum 
at early times.  The second possibility happens when the potential 
along the angular direction is flat enough during inflation.  In this
case $\Phi$ will not roll down to its minimum and can start at any
position at the onset of radial motion.  In both cases, the further 
$\Phi$ is away from the minimum, the larger the deflection of its 
trajectory and the wider the ellipsoidal shape will be.



\section{Complex Mathieu Resonance}

As described in the previous section, the potential for complex scalar
oscillations usually includes (as well as the scalar mass-squared  
terms) Hubble induced A-terms which are mainly effective during the  
first few cycles of oscillation. The A-terms provide a ``torque'' to 
the complex oscillation during the first few cycles, resulting in a 
net ``elliptic'' motion in the mass-term induced potential, after the 
A-terms have effectively ceased to be active. The resulting elliptic  
oscillation in the mass term potential will be damped by the Hubble  
drag, resulting in the ellipse shrinking over time.

In order to introduce new considerations characteristic of resonance 
with complex fields, without getting involved in model dependent 
details, in most of this paper we shall simply ignore the damping and
consider complex, elliptic, constant amplitude oscillations. In 
particular, this means we do not need to specify which particular 
field is considered (e.g. inflaton versus susy standard model flat 
direction), nor do we need to determine the cosmological details  
involved in determining expansion and damping at the time that the  
oscillations of the field in question occur. In addition to 
presenting a tractable and interesting mathematical problem, 
consideration of the undamped oscillation should also provide the 
essential features of the full cosmological case including the 
effects of expansion, at least in the generic case of broad-band 
resonance. This follows from the key observation of \cite{kls1,kls2} 
that in the broad-band case the resonant excitation of the decay 
product field occurs over a tiny fraction of the cycle of the driving 
field, when the latter passes near the origin, as only here do the 
decay product field dynamics depart from the adiabatic regime. So 
mode number excitation proceeds by a series of abrupt jumps, and the 
dynamics of a given jump may be considered with the instantaneous 
value of the oscillator parameters, resulting in the picture of 
``stochastic resonance'' analyzed in \cite{kls2}. The present paper 
discusses the changes in the dynamics of decay mode excitation which 
arise from the complex nature of the driving field oscillation---the 
differences in question arise from suppression of the adiabaticity  
violations that induce the jump in mode occupation numbers of the 
decay product field, so we expect that considerations using the 
instantaneous value of the driving oscillation amplitude should give 
insight in the complex case, much as such considerations did in the 
real stochastic resonance case. 

In any case, for the purposes of our present calculations we shall  
consider the parametric resonance production of decay product field  
modes $\Xi$, from phase-invariant coupling to constant-amplitude out 
of phase (``elliptic'') oscillations of a driving field $\Phi$. 
Detailed cosmological studies of applications to inflaton or moduli
oscillations will be considered elsewhere.

With the couplings discussed in the previous section, after the
A-terms cease to be effective, the equation of motion for the $\Xi$ 
field is of the form:
\begin{equation}
\label{eom}
   {\ddot{\Xi}}_{k} + 
      \left(
         \frac{k^{2}}{a^{2}}+m_{\Xi}^{2}+g^{2}|\Phi|^{2}
      \right)
      \Xi_{k}=0,
\end{equation}
where $\Xi_{k}$ is the decay product field mode with comoving  
wavenumber $k$, $a$ is the FRW scale factor, $m_{\Xi}$ the mechanical  
mass of the $\Xi$, and the superpotential coupling is as above. We 
note that both the real and imaginary piece of the $\Xi$ field will  
separately obey this equation, and hereafter we use $\chi$ to denote  
either the real or imaginary piece of $\Xi$. 

In our analysis, we will treat the physical momentum of the decay
field mode and the relative phase and amplitude of the driving field
oscillation as fixed parameters, and attempt to map out the regions 
of instability in their parameter ranges. As noted above, for the 
case of stochastic broad-band resonance, where the amplification 
occurs in small intervals while the field passes close to the origin,
one should be able to approximate the instantaneous behaviour within 
each interval by the corresponding behaviour of a system of the type 
we analyze here.

We decompose the driving field $\Phi$ into real and imaginary pieces 
as follows:
\begin{equation}
   \Phi = \phi_{R} + i\phi_{I}.
\end{equation}
By a phase rotation we put the largest amplitude component of  
oscillation into the real piece, and so we write:
\begin{eqnarray}
   \phi_{R} & = & \phi \sin(m_{\phi}t)  \\
   \phi_{I} & = & f\phi \cos(m_{\phi}t),
\end{eqnarray}
where now $\phi$ is the constant amplitude of the real component of
oscillation and $f \in [0,1]$ is the ``out of phase'' fractional 
component giving elliptic oscillation in the complex $\Phi$ plane; 
we will be particularly interested in the case where $f \ll 1$.

We wish to cast this into the canonical form of the (real) Mathieu  
equation:
\begin{equation}
   y^{\prime\prime} + (A - 2q\cos(2z))y = 0,
\end{equation}
where $'$ denotes derivative with respect to the independent variable
$z$. We begin by substituting our definition of the $\Phi$ field into
(\ref{eom}) above, giving
\begin{equation}
   {\ddot{\chi}}_{k} + 
      \left(
         \frac{k^{2}}{a^{2}}+m_{\chi}^{2}+g^{2}\phi^{2}
         \left(
	    \sin^{2}(m_{\phi}t)+f^{2}\cos^{2}(m_{\phi}t)
	 \right)
      \right)
      \chi_{k} 
      = 0.
\end{equation}
We replace $\cos^{2}(m_{\phi}t)$ with $1 - \sin^{2}(m_{\phi}t)$ and  
collect terms, giving
\begin{equation}
   {\ddot{\chi}}_{k} +
      \left(
         \frac{k^{2}}{a^{2}}+m_{\chi}^{2}+f^{2}g^{2}\phi^{2} + 
         \left(1-f^{2}\right)g^{2}\phi^{2}\sin^{2}(m_{\phi}t)
      \right)
      \chi_{k} 
      = 0.
\end{equation}
Using the half-angle formula 
$\sin^{2}\theta = \frac{1}{2}(1 - \cos2\theta)$ we obtain the form
\begin{equation}
   {\ddot{\chi}}_{k} + 
      \left(
         \frac{k^{2}}{a^{2}}+m_{\chi}^{2}+f^{2}g^{2}\phi^{2} + 
	 \frac{1}{2}\left(1-f^{2}\right)g^{2}\phi^{2}
	 \left(1-\cos(2m_{\phi}t)\right)
      \right)
      \chi_{k}
      = 0.
\end{equation}
which may be rewritten in the form of the Mathieu equation
\begin{equation}
   \chi_{k}^{\prime\prime} + 
      \left(A_{k}(f)-2q(f)\cos2z\right)\chi_{k} = 0
\end{equation}
with the following new identifications:
\begin{eqnarray}
   z & = & m_{\phi}t  \\
   A_{k}(f) 
      & = & \frac{\frac{k^{2}}{a^{2}}+m_{\chi}^{2}+f^{2}g^{2}\phi^{2}}
            {m_{\phi}^{2}} + 2q(f) 
	    \label{Akf}  \\
   q(f) 
      & = & \frac{(1-f^{2})g^{2}\phi^{2}}{4m_{\phi}^{2}}.
            \label{qf}
\end{eqnarray}
Notice that the coefficients of the Mathieu equation are now 
functions of the imaginary fraction $f$. This is an important 
feature, as it means that the characteristic behaviour of the 
parametric resonance as described by the Mathieu equation takes the 
same form in the complex case as in the real case; however, \emph{the 
relationship between the physical parameters of the process and the 
Mathieu coefficients is redefined}. 

So we see that there is a mapping that takes the Mathieu equation for 
the complex modes of the decay field when driven by the complex  
parametric field with out of phase real and imaginary pieces in its  
oscillation amplitude, and maps it to a real Mathieu equation for the 
oscillations of the real and imaginary pieces of the decay product  
field with shifted parameters. There are several features of this  
mapping that simply encapsulate physical features of the original  
problem.

First we note that for the case where $f = 0$, where the oscillation  
of the parametric driving field is along the real axis, the  
coefficients $A_{k}$ and $q$ reduce to those previously considered in 
the literature for the case of purely real parametric oscillation  
\cite{kls1,kls2}. In the other extreme limit, when $f = 1$, we have
$q(1) = 0$, meaning that one is restricted to be along the $A_{k}$ 
axis on the Mathieu equation stability diagram, allowing only 
non-resonant particle production. This corresponds to the physical  
observation that because the decay couplings were phase invariant, 
our original equation for the complex oscillations involved only the  
magnitude of $\Phi$ in the oscillation equation. In the case that the 
real and imaginary pieces of the $\Phi$ oscillation have the same  
amplitude ($f = 1$), then the coefficients in the $\chi$ equation of
motion become time independent and there can be no parametric
amplification. We also note that the imaginary fraction $f$ of the 
oscillations enters into the coefficients only as $f^{2}$, meaning 
that the effect is the same regardless of direction traveled around 
the ellipse (i.e. whether $f$ is positive or negative). This reflects
that fact that the original equation for the complex oscillations is 
second order and symmetric under time inversion, which corresponds to
having the parametric field circulate about the oval in the $\Phi$ 
plane in the opposite sense, or reversing the sign of $f$.      

Finally, in the intermediate regime $f \in (0,1)$, $q(f)$ is always  
lessened, while $A_{k}$ is always increased (compared to $f = 0$), so 
increasing $f$ never causes the system to leave the physical regime.
As noted above, increasing $f$ means moving ``inland'' on the 
stability diagram for the Mathieu equation. In general, this causes
suppression of the resonant growth of the $\chi$ modes; however, it 
also allows one to explain the counterintuitive observation that in 
certain cases the resonant band exponential growth parameter 
$\mu_{k}$ may actually increase as one turns on the out of phase 
component $f$. To understand this, imagine that for oscillations with
no imaginary fraction $f$ one is sitting in parameter space at the
lower border of one of the instability bands (where $\mu_{k}$ is 
zero). Now slowly increase $f$. The parameter mapping derived above 
implies that you start to move in a ``northwest'' direction on the 
band chart into the instability band. As such, your $\mu_{k}$ begins 
to increase. As you continue to increase $f$, you will eventually hit
the maximum possible $\mu_{k}$ along your trajectory, after which 
your $\mu_{k}$ begins to drop again. Eventually you will leave the 
instability band altogether for sufficiently large $f$. For very high
order instability bands, it should be possible to encounter many 
bands along one trajectory of increasing $f$.

From a different point of view, were one to look at the instability  
diagram as a function of the the $A_{k}$ and $q$ of standard real  
Mathieu parametric resonance (ie. as a function of our $A_{k}(0)$ and
$q(0)$, for different values of $f$), the effect of turning on $f$  
would be seen to manifest itself as both a narrowing and a downwards 
shift of the instability bands as $f$ increased. In addition, 
isocontours of $\mu_{k}$ would be seen to ``flow out'' of the bands 
as $f$ increases. For $f=1$ each instability band collapses to a line
with $\mu_{k} = 0$, as with a phase independent coupling of the 
parameter field to the mode field there would be no time dependence 
in the mode field equation of motion, and the system would be stable 
for all $A_{k}$ and $q$. Figure 1 illustrates in detail both the 
bending of the bands and the decrease of $\mu_{k}$ for the first 
resonance band as we turn on the imaginary fraction $f$ of our 
oscillations. We expect that the suppression of resonance for fixed 
non-zero imaginary fraction $f$ is stronger in higher resonance bands,
as at larger $q$ the resonance proceeds by violation of adiabaticity 
in the decay mode evolution, and for large $q$ at fixed $f$ the 
induced decay mode field mass is always large. This is illustrated in
Figure 2 where we show the quenching of resonance for the first 7
resonance bands, as $f$ is turned on. We see that there is more
severe suppression for the higher bands, in accord with our physical
intuition; in the next section we will analytically estimate the
extent of the domain of surviving resonant bands, for fixed imaginary
fraction $f$.



\section{Parameter Domain For Broad-Band Resonance} 

Let us first briefly discuss the effect of mixing in the narrow-band  
regime.  The narrow-band resonance in the case of a real oscillating  
field is efficient for 
${( {m \over g})}^{{4 \over 3}}\leq \phi \leq {m \over g}$ 
\cite{stb}.  In the case of a complex field with mixing parameter $f$
this reads as
\beq
   \left(
      \frac{{m}^{2} + {f}^{2}{g}^{2}{\phi}^{2}}{(1-{f}^{2}){g}^{2}}
   \right)^\frac{2}{3}  
   \leq \phi \leq 
   \left(
      \frac{{m}^{2} + {f}^{2}{g}^{2}{\phi}^{2}}{(1-{f}^{2}){g}^{2}}
   \right)^\frac{1}{2},
\eeq
where $m$ and $g$ are replaced with 
$\sqrt{{m}^{2} + {f}^{2}{g}^{2}{\phi}^{2}}$ and $\sqrt{1-{f}^{2}}g$, 
respectively.  It is easily seen that the condition for an efficient 
narrow-band resonance remains almost unchanged in the complex case, 
unless $f$ is close to 1. Therefore, in physically interesting 
situations the narrow-band resonance will not be substantially 
affected by the mixing. Of course, in the extreme case with $f=1$,
there is no time variation in the Mathieu equation and, hence, 
no narrow-band resonance.     

In the case of real broad-band parametric resonance, Kofman, Linde,
and Starobinski \cite{kls1,kls2} argue that the requirement of 
adiabaticity violation for broad-band parametric amplification
implies that it only occurs with significant $\mu_{k}$ for 
$A_{k} - 2q \lesssim \sqrt{q}$. Their argument presupposes that 
the decay terminates after it has entered an ``explosive'' phase, 
where the effective mass of the decaying $\phi$ is dominated by the 
coupling to the plasma of decay product $\chi$ modes which has been 
built up by parametric resonance decay. The effective physical 
3-momentum of the quasi-relativistic decay $\chi$ modes is of order 
their energy, which is no more than of order the induced mass of the 
decaying $\phi$; this is of order $g \chi_{\mathrm{end}}$ which in 
turn is of order $g \phi_{\mathrm{end}}$ which can also be written  
$\sqrt{g {m^{\mathrm{eff}}_{\phi}}\phi_{\mathrm{end}}}$. So 
$(k_{\mathrm{phys}}^{2} / m^{2}_{\phi}) \lesssim 
({g \phi_{\mathrm{end}}}/ m^{\mathrm{eff}}_{\phi})$, which can be 
rewritten as $A_{k} - 2q \lesssim \sqrt{q}$. For a detailed
discussion we refer to the treatment in \cite{kls2}. 

We have seen in the preceding section that the case of complex  
oscillation with imaginary fraction $f$ can be mapped onto a Mathieu  
equation with shifted resonance parameters.  By substituting the  
``shifted'' parameters induced by the non-zero imaginary component of
oscillation, we should thus be able to determine what range of $q$  
supports broad-band resonance for oscillation with a given fraction of
out of phase imaginary component for the oscillation of the driving  
parameter. 

Recall the expressions for the equivalent shifted $A_{k}(f)$  
and $q(f)$ from equations (\ref{Akf}) and (\ref{qf}) respectively. 
Substituting these expressions into the relation 
$A_{k}-2q \lesssim \sqrt{q}$ allows us to write it as:
\beq
   \frac{\frac{k^{2}}{a^{2}}+m_{\chi}^{2}+f^{2}g^{2}\phi^{2}}
      {m_{\phi}^{2}} 
   \lesssim
   \frac{{\sqrt{(1-f^{2})}}g\phi}{2m_{\phi}}. 
\eeq
This leads us to the relation
\beq
    A_{k}(0)-2q(0) + 4f^{2}q(0) 
    \lesssim
    \sqrt{1-f^{2}}\sqrt{q(0)}, 
\eeq
or, defining $E_{k} \equiv A_{k}(0)-2q(0)$, we have 
\beq
   E_{k} + 4f^{2}q(0) 
   \lesssim
   \sqrt{1-f^{2}}\sqrt{q(0)} 
   \label{fcrit}.
\eeq
If we recall that physical values of $E_{k}$ are positive 
semi-definite, we find that for a fixed non-zero imaginary fraction 
$f$ there is an upper bound on the parameter $q(0)$ for which 
resonance occurs, and the allowed range of resonant $q(0)$ is bounded
above as $\frac{1-f^{2}}{16 f^{4}}$. (For the small imaginary 
fractions $f$ of physical interest, however, the weaker approximate
bound of $\frac{1}{16} f^{-4}$ will suffice). So instead of an an
ever widening resonance region above the $A_{k} = 2q$ line, with 
thickness of order $\sqrt{q}$, as one has in the real case, in the 
complex case with fixed non-zero imaginary fraction $f$, one instead 
has a region above the  $A_{k} = 2q$ line of finite extent, with
an upper bound on the values of the $q$ parameter which can result in
resonance. This is qualitatively reasonable, as a fixed imaginary  
fraction $f$ for the oscillation means that as we scale up $q$ the  
ellipse of $\Phi$ broadens as it lengthens, preserving its shape; so  
throughout the $\Phi$ oscillation $|\Phi|$ has a large value, 
inducing a large mass for the modes of the decay field $\Xi$, which 
in turn leads to adiabatic evolution of the $\Xi$, and suppression of
broad-band parametric resonance production of the $\Xi$.



\section{Complex Resonance in ``Instant Preheat''}

Recently, a simpler method of efficient scalar field decay has been  
proposed, called ``instant preheat'' \cite{fkl}. Within models of 
this type the decaying field rolls once through the origin, at which 
point the mass of the decay product field to which it is coupled
passes through zero, and modes of the decay product field experience  
non-adiabatic excitation. As the decaying field rolls away from zero  
(perhaps monotonically) the modes of the decay product field grow in  
mass; they drain energy from the decaying field through their mass. As
the mass of the modes of the decay product field grows, so does their 
decay width; their subsequent decay, after their mass and decay width 
have grown sufficiently, then releases the energy they have taken from
the original decay field, and dumps it into their final decay 
products, which thermalize the resulting energy.

It is interesting to note that while final state effects such as 
rescattering, backreaction, or plasma masses can prevent preheating 
from occurring, they are unimportant in the instant preheating
scenario. The reason is that for these effects to become important, 
(at least) several oscillations are needed to build up a large enough
occupation number for the final state field.  In the instant
preheating, on the other hand, the energy drain from the oscillating
field occurs during each half of an oscillation.  In fact, instant
preheating can be efficient even if the adiabaticity condition is
violated only during the first half of the first oscillation. 
Therefore, instant preheating is essentially unaffected by the final
state effects.  The mixing of the real and imaginary parts of the  
oscillating field, on the other hand, has the same effect in the  
instant preheating case as in the standard preheating scenario.  We 
recall that the torque from A-terms deflects the trajectory of the 
oscillating field from that of a straight line.  The Hubble induced 
A-terms have their largest value at the beginning of the oscillations,
and rapidly decrease with Hubble expansion.  This means that the
deflection is largest during the initial oscillations.   Thus, a
large enough $f$ to restore adiabaticity in the preheating case
could do the same for the instant preheating case.    



\section{Non-Convex Potentials}

Another possibility to achieve rapid decay of a homogeneous condensate  
occurs in the case that the potential governing the evolution of the  
condensate scalar has non-convex behaviour over some region of field  
space \cite{gpr,ks}.
In this circumstance, it becomes energetically favorable for a scalar 
condensate in the non-convex region to decompose into inhomogeneous  
modes; provided the inhomogeneity occurs over long enough wavelengths,
the price one pays in kinetic energy for the inhomogeneity is more
than compensated by the decreased average potential energy of the
regions of field excess and deficit compared to the average field
value. This produces a wavenumber band for exponential growth of the
mode amplitudes. This has been considered in both the case of
inflaton decay \cite{gpr}, and in the case of the growth of
inhomogeneities in scalar condensates corresponding to F-flat and
D-flat directions of the standard model with non-convex potentials
of the type arising from gauge-mediated supersymmetry breaking
\cite{ks}. It is clear that this is one type of instability which is
not vitiated by having the scalar order parameter complex or
involving multiple scalar fields. If there is a region in field space
with respect to which the potential is non-convex in some direction,
then fluctuations corresponding to modes of the field variation in
that field direction, of sufficiently long wavelength, will win on the
potential versus kinetic energy budget, and grow exponentially.
Indeed the treatment of (complex) flat directions in the
supersymmetric standard model in \cite{ks} explicitly analyzes the
conditions for instability of a complex field with a potential which
is a non-convex function of the field modulus, and exhibits the 
resulting instability bands.



\section{Other Couplings}

Here, we briefly comment on the situation for another type of coupling  
between $\Phi$ and $\Xi$ fields which is also of interest and  
application.  This is the coupling 
${g}^{2}{({\phi}_{R}{\chi}_{R} + {\phi}_{I}{\chi}_{I})}^{2}$, 
where its simplest manifestation is for the potential 
$V(\Phi) = \frac{1}{4} \lambda {| \Phi |}^{4}$, with $\Phi$ and 
$\Xi$ being the same field. It also arises in supersymmetric models 
from the D-term part of the scalar potential. This type of coupling 
leads to the mixing of ${\chi}_{R}$ and ${\chi}_{I}$ mode equations:
\begin{eqnarray}
   {\ddot{\chi}}_{R,k} + 
   \left(
       {{k}^{2} \over {a}^{2}} 
       + {{m}_{\chi}}^{2}{g}^{2}{{\phi}_{R}}^{2}
   \right)
   {\chi}_{R,k} + {g}^{2}{\phi}_{R}{\phi}_{I}{\chi}_{I,k} 
   & = & 0   \\
   {\ddot{\chi}}_{I,k} + 
   \left(
      {{k}^{2} \over {a}^{2}} 
      +{{m}_{\chi}}^{2} {g}^{2}{{\phi}_{I}}^{2}
   \right)
   {\chi}_{I,k} + {g}^{2}{\phi}_{R}{\phi}_{I}{\chi}_{R,k} 
   & = & 0.
\end{eqnarray}
In this case the mass eigenstates are 
${{\phi }_{R}{\chi}_{R} + {\phi}_{I}{\chi}_{I}\over \phi}$ 
and ${{\phi }_{I}{\chi}_{R} - {\phi}_{R}{\chi}_{I}\over \phi}$ 
instead of ${\phi}_{R}$ and ${\phi}_{I}$ themselves.  For 
oscillatory motion of ${\phi}_{R}$ and ${\phi}_{I}$ with a phase 
difference, there are two periodic changes that may lead to 
resonance: change in the mass eigenvalues (the usual parametric 
resonance) and change in the mass eigenstates.  They can't be simply 
superimposed and it is not very easy to give rough arguments for the 
instability bands and the respective value of $\mu_{k}$'s.  The
important point is that for such a coupling, even in the $f=1$ case
there is still time variation in mode equations.  This variation is
present in both the mass eigenstates and mass eigenvalues.    



\section{Cosmic Expansion and Complex Resonance}

So far, we have considered modifications to parametric resonance decay
which arise in complex field oscillations in the absence of effects of
Hubble expansion. As we noted above, since broad-band resonance is 
induced by non-adiabaticity of the $\chi$ evolution during small 
intervals of the $\phi$ oscillation, instantaneous approximation of
the $\chi$ excitation should be a useful guide during each of the
jumps in mode number. Cosmic expansion then functions to shift the
parameters of the oscillator between episodes of mode excitation as
$\phi$ passes near zero. We now examine the implications of this in
both the narrow- and broad-band cases.

Implications for the narrow-band case are simple; as we have seen, the
introduction of a phase difference between real and imaginary
components of our complex inflaton field $\Phi$ only kills narrow-band
resonance for phase differences approaching $\frac{\pi}{2}$, or, in
the language of this paper, for $f \cong 1$. Therefore, the resonance
should be qualitatively the same in the static approximation and with
the Hubble expansion included.

For broad-band resonance the situation is completely different.
According to equation (\ref{fcrit}), the broad-band resonance is
shut off for $q > \frac{1}{16}f^{-4}$.  In the static limit $f$ and 
$q$ are both constant and resonance is either suppressed, or viable.
In an expanding universe, $f$ eventually becomes approximately
constant as the Hubble induced A-terms turn off (indeed, as pointed
out earlier, after several Hubble times the motions along the real
and imaginary directions are decoupled and free), while, on the other
hand, $q(t)={({g \phi(t) \over 2m})}^{2}$ is redshifted as ${a}^{-3}$.
This implies that even if the resonance is suppressed initially, it
may be initiated after a sufficient time such that 
$q(t) < {1 \over 16} f^{-4}$.  The right-hand side is less than 1 (or
very close to it) for $f \gtrsim {1 \over 2}$.  Therefore, in the
case of large out of phase components of oscillation, the broad-band
resonance is killed and resonance may resume only in the narrow-band
regime at a later time.  In most physically interesting cases,
however, $f < {1 \over 2}$ and the right-hand side is considerably 
greater than 1.  In such cases, broad-band resonance is not 
eliminated in an expanding universe, but rather delayed.  The 
parameter $f$ is determined by the action of the scalar potential 
(including A-terms) for the oscillating field, and after the initial
oscillations it often becomes effectively time-independent.
Depending on the dimensionality of the A-term, it may be a function
of ${q}_{i}$, the value of $q$ at the start of oscillations.  If 
${q}_{i} < {1 \over 16} f^{-4}$, the onset of broad-band resonance
will be unaffected.  For ${q}_{i} > {1 \over 16} f^{-4}$, resonance
is delayed initially, but will resume after sufficient expansion such
that $q < {q}_{\mathrm{eq}} = {1 \over 16} f^{-4}$.  A larger $f$
leads to a smaller ${q}_{\mathrm{eq}}$, for $f \gtrsim {1 \over 2}$ 
we have ${q}_{\mathrm{eq}} < 1$ and resonance can only occur in the
narrow-band regime.  For $f = 1$ resonance is eliminated.  

Even though the broad-band resonance (for interesting cases) is only
delayed in an expanding universe, the mixing still has important 
consequences. Perhaps the most notable example relates to the
production of superheavy particles during resonance.  In the standard
preheating, $\Xi$'s with a mass up to ${q}^{ {1 \over 4}}{m}_{\phi}$
can be produced.  A reduction in $q$ at the onset of resonance
implies a reduction in the maximum mass of produced particles. This
is even more pronounced in the instant preheating case.  Here $\Xi$
decay products $\Psi$ with masses up to ${q}^{1 \over 2}{m}_{\phi}$
and which have a large enough coupling $h$ to $\Xi$, can be produced
\cite{fkl}.  A smaller ${q}_{\mathrm{eq}}$ has a two-fold effect in 
this case.  First, the allowable masses are smaller, and second, the 
decay rate ${\Gamma}_{d} = {{h}^{2} \over 8 \pi} g \phi$ may not be
large enough (compared to the frequency of oscillations ${m}_{\phi}$)
for efficient production of $\Psi$'s.  It is easily seen that 
${\Gamma}_{d} \leq {m}_{\phi}$ for 
$h \lesssim 4 {\pi}^{{1 \over 2}} f$. Therefore, $\Psi$ production is
not efficient if $h \lesssim 4 {\pi}^{{1 \over 2}} f$. Even for $h 
\gg  4 {\pi}^{{1 \over 2}} f$, only $\Psi$'s with a mass ${m}_{\psi} 
\lesssim {1 \over 4{f}^{2}} {m}_{\phi}$ can be produced.



\section{Conclusions}

In this paper, we have considered the changes to the standard picture
of parametric resonance decay of a real homogeneous cosmological
scalar field which arise if the field is instead complex, with out of
phase oscillation of its real and imaginary components and a phase
invariant decay coupling. For the case of complex Mathieu type
resonance, we give an explicit mapping to a corresponding real Mathieu
resonance with shifted parameters that encode the effects of the out
of phase components of the oscillating decay field. We showed the
resulting effects on the instability bands, demonstrating how they
shift and shrink with increasing out of phase (``elliptic'') component
of the driving field motion, limiting the swath of instability to a
finite area on the $A_{k}$-$q$ chart, and eliminating broad-band
resonance in the higher modes. We argued that similar effects may be
present in the case of complex field models of ``instant preheat,''
but that instabilities due to regions in field space with non-convex
potentials are qualitatively the same in the complex case. Finally, in
the context of an expanding FRW universe, we noted that the presence
of a fraction of out of phase oscillation would usually delay the
onset of parametric resonance, but not eliminate it entirely.

\vskip .1in


\noindent{ {\bf Acknowledgements} }

\noindent We would like to thank Andrei Linde for very helpful 
discussions, and explanations of broad-band resonance in the real 
case. This work was supported in part by the Natural Sciences and 
Engineering Research Council of Canada. RA would like to thank the 
CERN Theory Division for kind hospitality during part of this 
research.

\newpage

\begin{center}
\textbf{Figure Captions}
\end{center}

\vskip .1in

\begin{figure}[!ht]
\caption{The Mathieu equation stability diagram for the first 
resonance band. The physical region lies above and to the left of 
the diagonal line $A_{k} = 2q$. Contour lines represent isocontours 
of $\mu$ starting from $\mu = 0$ (band boundary) and increasing by 
units of $0.1$ as one moves from left to right in the band. Different 
panels represent different imaginary fractions $f$: (a) $f = 0$; 
(b) $f = 0.2$; (c) $f = 0.4$; (d) $f = 0.6$.}
\end{figure}

\begin{figure}[!ht]
\caption{The Mathieu equation stability diagram for the first 7 
resonance bands. The physical region lies above and to the left of  
the diagonal line $A_{k} = 2q$. Contour lines represent isocontours 
of $\mu$ starting from $\mu = 0$ (band boundary) and increasing by 
units of $0.1$ as one moves from left to right in the bands.
Different panels represent different imaginary fractions $f$: 
(a) $f = 0$; (b) $f = 0.2$; (c) $f = 0.4$; (d) $f = 0.6$.}
\end{figure}

\begin{center}
   \resizebox{!}{\textheight}{\includegraphics{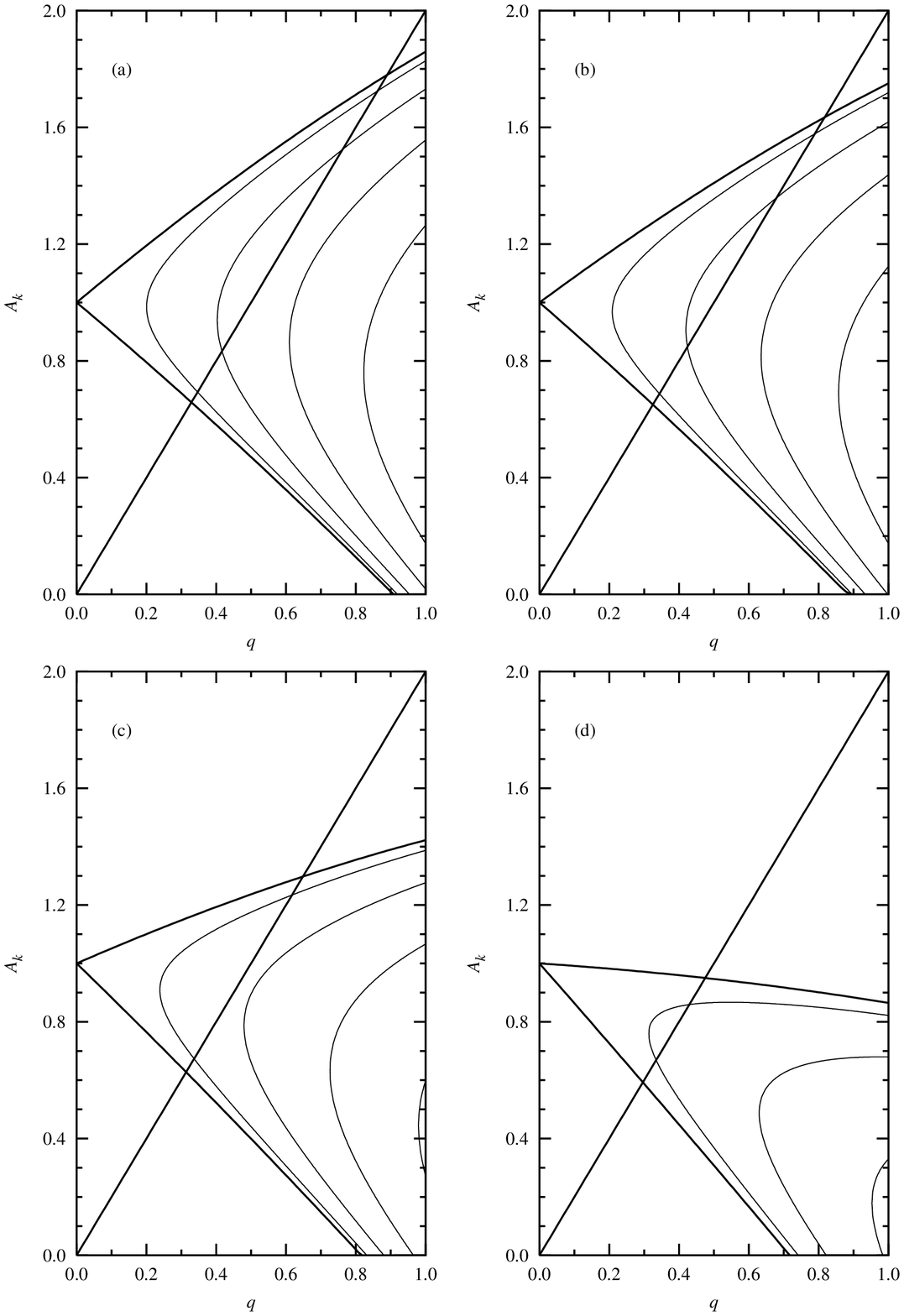}}
\end{center}

\begin{center}
   \resizebox{!}{\textheight}{\includegraphics{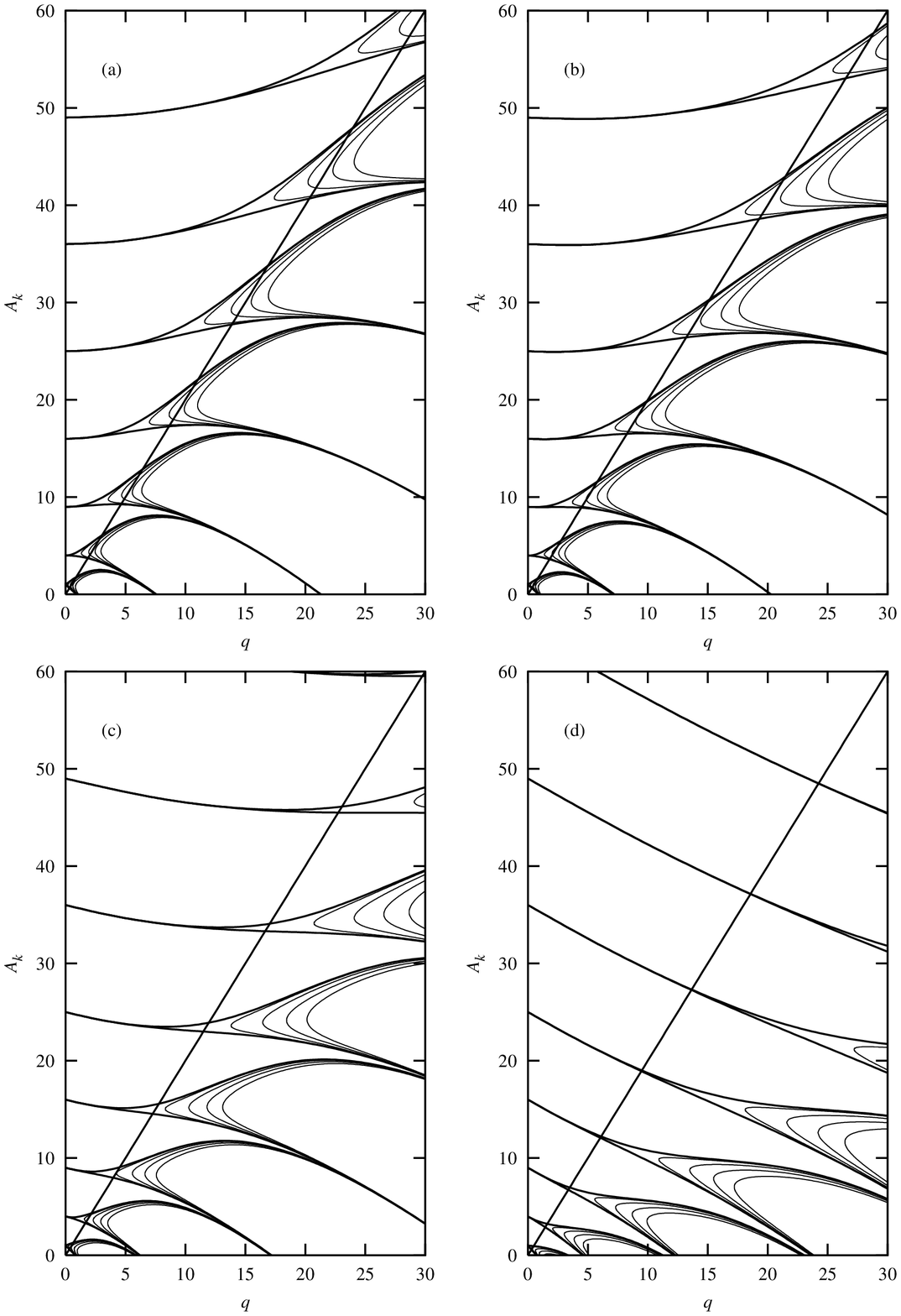}}
\end{center}


\newpage


\begin{thebibliography}{99}
\bibitem{infl} for reviews see: A. D. Linde, \underline{Particle 
Physics And Inflationary Cosmology} Harwood (1990);
K. A. Olive, Phys.Rep. {\bf C190}(1990), 307.

\bibitem{sr} A. Dolgov and A. Linde, 
Phys. Lett. {\bf 116B} (1982), 329;
L. F. Abbott, E. Farhi, and M. Wise, 
Phys. Lett. {\bf 117B} (1982), 29.

\bibitem{ekn} J. Ellis, J. E. Kim, and D. V. Nanopolous. 
Phys. Lett. {\bf B145} (1984), 181;
M. Kawasaki and T. Moroi, Prog. Theor. Phys. {\bf 93} (1995), 879;
for review see: S. Sarkar, Rept. Prog. Phys. {\bf 59} (1996), 1493. 
  
\bibitem{kls1}  L. Kofman, A. D. Linde, and A. A. Starobinski, 
Phys. Rev. Lett. {\bf 73} (1994), 3195.

\bibitem{kls2} L. Kofman, A. D. Linde, and A. A. Starobinski, 
Phys. Rev. {\bf D56} (1997), 3258.

\bibitem{stb} Y. Shtanov, J. Traschen, and R. Brandenberger, 
Phys. Rev. {\bf D51} (1995), 5438.

\bibitem{k1} D. I. Kaiser, Phys. Rev. {\bf D 53} (1996) 1776; 
Phys. Rev. {\bf D56} (1997), 706.

\bibitem{kt1} S. Yu. Khlebnikov and I. I. Tkachev, 
Phys. Rev. Lett {\bf 77} (1996), 219;
S. Yu. Khlebnikov and I. I. Tkachev, 
Phys. Lett. {\bf B390} (1997), 80. (1997).

\bibitem{kt2} S. Yu. Khlebnikov and I. I. Tkachev, 
Phys. Rev. Lett {\bf 79} (1997), 1607.

\bibitem{pr} T. Prokopec and T. Roos,
 Phys. Rev. {\bf D55} (1997) 169.
 
\bibitem{gkls/k2} P. B. Green, L. Kofman, A. D. Linde, 
and A. A. Starobinski, Phys. Rev. {\bf D56} (1997), 6175;
D. I. Kaiser, Phys. Rev. {\bf D57} (1998), 702.

\bibitem{gpr} B. Greene, T. Prokopec, and T. Roos, 
Phys. Rev. {\bf D56} (1997), 6484.

\bibitem{zhs} I. Zlatev, G. Heuy, and P. J. Steinhardt, 
Phys. Rev. {\bf D57} (1998), 2152.

\bibitem{gk} P. B. Green and L. Kofman, 
Phys. Lett. {\bf B44} (1999), 6.

\bibitem{gks} P. B. Green, L. Kofman, and A. A. Starobinski, 
Nucl. Phys. {\bf B543} (1999), 423.

\bibitem{fkl} G. Felder, L. Kofman, and A. D. Linde, 
Phys. Rev. {\bf D59} (1999), 123523;
G. Felder, L. Kofman, and A. D. Linde, hep-ph/9903350.

\bibitem{nsr} L. Kofman, A. D. Linde, and A. A. Starobinski, 
Phys. Rev. Lett. {\bf 76} (1996), 1011;
I. I. Tkachev, Phys. Lett. {\bf B376} (1996), 35.

\bibitem{tf}  S. Yu. Khlebnikov, L. Kofman, A. D. Linde, 
and I. I. Tkachev, Phys. Rev. Lett. {\bf 81} (1998), 2012;
I. I. Tkachev, S. Yu. Khlebnikov, L. Kofman, and A. D. Linde, 
Phys. Lett. {\bf B440} (1998), 262;
S. Kasuya and M. Kawasaki, Phys. Rev. {\bf D58} (1998), 083516.

\bibitem{gb1} E. W. Kolb, A. D. Linde, and A. Riotto, 
Phys. Rev. Lett. {\bf 77} (1996), 4290.

\bibitem{gb2} E. W. Kolb, A. Riotto, and I. I. Tkachev, 
Phys. Lett. {\bf B423} (1998), 348.

\bibitem{sb}  G. Anderson, A. D. Linde, and A. Riotto, 
Phys. Rev. Lett {\bf 77} (1996), 3716; 
G. Dvali and A. Riotto, Phs. Lett. {\bf B388} (1996), 247. 

\bibitem{spp} D. J. H. Chung, E. W. Kolb, and A. Riotto, 
Phys. Rev. Lett ,{\bf81} (1998), 4048; 
Phys. Rev. {\bf D59} (1999), 023501.   
  
\bibitem{gp1} H. Fujisaki, K. Kumekawa, M. Yamguchi, 
and M. Yoshimura, Phys. Rev. {\bf D54} (1996), 2494.

\bibitem{gp2} A. L. Maroto and A. Mazumdar, hep-ph/9904206; 
R. Kallosh, L. Kofman, A. D. Linde, and A. Van Proeyen, 
hep-th/9907124; G. Giudice, I. I. Tkachev, and A. Riotto, 
hep-ph/9907510.

\bibitem{kk} S. Kasuya and M. Kawasaki, 
Phys. Lett. {\bf B388} (1996), 686. 

\bibitem{ac} R. Allahverdi and B. A. Campbell, 
Phys. Lett. {\bf B395} (1997), 169.  

\bibitem{drt} M. Dine, L. Randall, and S. Thomas, 
Nucl. Phys. {\bf B458} (1996), 291.

\bibitem{gmo} M. K. Gaillard, H. Murayama, and K. A. Olive, Phys.
Lett. {\bf B355} (1995), 71.

\bibitem{ks} A. Kusenko and M. Shaposhnikov, 
Phys. Lett. {\bf B418} (1998), 46. 


\end{thebibliography}
\end{document}